\documentclass[12pt]{article}
\usepackage{amsfonts}
\usepackage{amssymb}
\begin{document}
\vspace{-48pt}
\title{{\bf On a possible quantum contribution \\ to the red shift}\footnote{Talk prepared for the Conference {\it{Problems of Practical Cosmology}}, Saint Petersburg, Russia, June 23 -- 27, 2008}}
\author{K. Urbanowski\footnote {e--mail:
K.Urbanowski@proton.if.uz.zgora.pl; K.Urbanowski@if.uz.zgora.pl}\\
University of Zielona G\'{o}ra, Institute of Physics, \\
ul. Prof. Z. Szafrana 4a, 65--516 Zielona G\'{o}ra, Poland.}
\maketitle
\vspace*{-12pt}
\begin{abstract}
We consider an effect generated by the nonexponential behavior
of the survival amplitude of an unstable state in the long time region:
In 1957 Khalfin proved that this  amplitude
tends to zero
as $t$ goes to the infinity more slowly than any exponential function of $t$.  This
effect can be described in terms of time-dependent decay rate $\gamma(t)$ and
then the Khalfin result means that this $\gamma(t)$ is not a constant for long
times but that it tends to zero as $t$ goes to the infinity. It
appears that a similar
conclusion can be drawn for the energy of the unstable state for a large class
of models of unstable particles: This energy should be much smaller for suitably
long times $t$ than the energy of this state for $t$ of the order of the lifetime of
the considered state. Within a given model we show that the energy corrections  in the long ($t
\rightarrow \infty$) and relatively short (lifetime of the state)
time regions, are different. It is shown that  these corrections decrease to
${\cal E} = {\cal E}_{min} < {\cal E}_{\phi}$
as $t \rightarrow \infty$, where ${\cal E}_{\phi}$ is the energy
of the system in the state $|\phi \rangle$
measured at times $t \sim \tau_{\phi}= \frac{\hbar}{\gamma}$.
This is a purely quantum mechanical effect.
It is hypothesized that there is a possibility to detect this effect by
analyzing the spectra of distant astrophysical objects.
The above property of unstable states may influence the measured
values of astrophysical and cosmological parameters.
\end{abstract} \pagebreak

\section{Introduction}
Within the quantum theory the state vector at time $t$, $|\Phi
(t)\rangle$, for the physical system under consideration which
initially (at $t = t_{0} =0$) was in the state $|\Phi\rangle$   can
be found  by solving the  Sch\"{o}dinger equation
\begin{equation}
i\hbar \frac{\partial}{\partial t} |\Phi (t) \rangle = H |\Phi
(t)\rangle, \;\;\;\;\; |\Phi (0) \rangle = |\Phi\rangle, \label
{Schrod}
\end{equation}
where $|\Phi (t) \rangle, |\Phi \rangle \in {\cal H}$,  ${\cal H}$
is the Hilbert space of states of the considered system, $\| \,|\Phi
(t) \rangle \| = \| \,|\Phi \rangle \| = 1$ and $H$ denotes the
total selfadjoint Hamiltonian for the system. If one considers an
unstable state $|\Phi \rangle \equiv |\phi\rangle$ of the system
then using the solution $|\phi (t)\rangle$ of Eq. (\ref{Schrod}) for
the initial condition $|\phi (0) \rangle = |\phi\rangle$ one can
determine the decay law, ${\cal P}_{\phi}(t)$ of this state decaying
in vacuum
\begin{equation}
{\cal P}_{\phi}(t) = |a(t)|^{2}, \label{P(t)}
\end{equation}
where $a(t)$ is  probability amplitude of finding the system at the
time $t$ in the initial state $|\phi\rangle$ prepared at time $t_{0}
= 0$,
\begin{equation}
a(t) = \langle \phi|\phi (t) \rangle . \label{a(t)}
\end{equation}
We have
\begin{equation}
a(0) = 1. \label{a(0)}
\end{equation}
From basic principles of quantum theory it is known that the
amplitude $a(t)$, and thus the decay law ${\cal P}_{\phi}(t)$ of the
unstable state $|\phi\rangle$, are completely determined by the
density of the energy distribution $\omega({\cal E})$ for the system
in this state \cite{Fock},
\begin{equation}
a(t) = \int_{Spec.(H)} \omega({\cal E})\;
e^{\textstyle{-\frac{i}{\hbar}\,{\cal E}\,t}}\,d{\cal E}.
\label{a-spec}
\end{equation}
where $\omega({\cal E}) > 0$.

Note that (\ref{a-spec}) and (\ref{a(0)}) mean that there must be
\begin{equation}
a(0) = \int_{Spec. (H)} \omega ({\cal E})\,d{\cal E} = 1.
\label{a(0)-spec}
\end{equation}
From the last property  and from the Riemann--Lebesgue Lemma it
follows that the amplitude $a(t)$, being the Fourier transform of
$\omega ({\cal E})$ (see (\ref{a-spec})),  must tend to zero as $t
\rightarrow \infty$ \cite{Fock}.

In \cite{Khalfin} assuming that the spectrum of $H$ must be bounded
from below, $(Spec.(H)\; > \; -\infty)$, and using the Paley--Wiener
Theorem \cite{Paley} it was proved that in the case of unstable
states there must be
\begin{equation}
|a(t)| \; \geq \; A\,e^{\textstyle - b \,t^{q}}, \label{|a(t)|-as}
\end{equation}
for $|t| \rightarrow \infty$. Here $A > 0,\,b> 0$ and $ 0 < q < 1$.
This means that the decay law ${\cal P}_{\phi}(t)$ of unstable
states decaying in the vacuum, (\ref{P(t)}), can not be described by
an exponential function of time $t$ if time $t$ is suitably long, $t
\rightarrow \infty$, and that for these lengths of time ${\cal
P}_{\phi}(t)$ tends to zero as $t \rightarrow \infty$  more slowly
than any exponential function of $t$. The analysis of the models of
the decay processes shows that ${\cal P}_{\phi}(t) \simeq
e^{\textstyle{- \frac{\gamma_{\phi}^{0} t}{\hbar}}}$, (where
$\gamma_{\phi}^{0}$ is the decay rate of the state $|\phi \rangle$),
to an very high accuracy  for a wide time range $t$: From $t$
suitably later than some $T_{0} \simeq t_{0}= 0$ but $T_{0} >
t_{0}$ up to $t \gg \tau_{\phi} = \frac{\gamma_{\phi}^{0}}{\hbar}$
and smaller than $t = t_{as}$, where $t_{as}$ denotes the
time $t$ for which the nonexponential deviations of $a(t)$
begin to dominate (see eg., \cite{Khalfin}, \cite{Goldberger} --
\cite{Greenland}). From this analysis it follows that in the general
case the decay law ${\cal P}_{\phi}(t)$ takes the inverse
power--like form $t^{- \lambda}$, (where $\lambda
> 0$), for suitably large $t \geq t_{as}\gg \tau_{\phi}$
\cite{Khalfin}, \cite{Goldberger} -- \cite{Peres}. This effect is in
agreement with the  general result (\ref{|a(t)|-as}).  Effects of this type
are sometimes called the "Khalfin effect" (see eg.
\cite{Arbo}).

The problem how to detect possible deviations from the exponential form of
${\cal P}_{\phi}(t)$ at the long time region has been attracting the  attention of physicists since the
first theoretical predictions of such an effect \cite{Wessner,Norman1,Greenland}.
Many tests of the decay law performed some time ago
did not indicate
any deviations
from the exponential form of ${\cal P}_{\phi}(t)$ at  the
long time region.
Nevertheless conditions leading to the nonexponetial behavior
of the amplitude $a(t)$ at long times were studied theoretically \cite{seke} -- \cite{jiitoh}.
Conclusions following from these studies were applied successfully in experiment described  in \cite{rothe},
where the experimental evidence of deviations from the exponential decay law at long times was
reported.
This result gives rise to another problem which now becomes important: if (and how) deviations from the
exponential decay law at long times affect the energy of the unstable state
and its decay rate at this time region.

Note that in fact the amplitude $a(t)$ contains information about
the decay law ${\cal P}_{\phi}(t)$ of the state $|\phi\rangle$, that
is about the decay rate $\gamma_{\phi}^{0}$ of this state, as well
as the energy ${\cal E}_{\phi}^{0}$ of the system in this state.
This information can be extracted from $a(t)$. Indeed if
$|\phi\rangle$ is an unstable (a quasi--stationary) state then
\begin{equation}
a(t)  \cong e^{\textstyle{ - \frac{i}{\hbar}({\cal E}_{\phi}^{0} -
\frac{i}{2} \gamma_{\phi}^{0})\,t }}. \label{a-q-stat}
\end{equation}
So, there is
\begin{equation}
{\cal E}_{\phi}^{0} - \frac{i}{2} \gamma_{\phi}^{0} \equiv i
\hbar\,\frac{\partial a(t)}{\partial t} \; \frac{1}{a(t)},
\label{E-iG}
\end{equation}
in the case of quasi--stationary states.

The standard interpretation and understanding of the quantum theory
and the related construction of our measuring devices are such that
detecting the energy ${\cal E}_{\phi}^{0}$ and decay rate
$\gamma_{\phi}^{0}$ one is sure that the amplitude $a(t)$ has the
form (\ref{a-q-stat}) and thus that the relation (\ref{E-iG})
occurs. Taking the above into account one can define the "effective
Hamiltonian", $h_{\phi}$, for the one--dimensional subspace of
states ${\cal H}_{||}$ spanned by the normalized vector
$|\phi\rangle$ as follows (see, eg. \cite{PRA})
\begin{equation}
h_{\phi} \stackrel{\rm def}{=}  i \hbar\, \frac{\partial
a(t)}{\partial t} \; \frac{1}{a(t)}. \label{h}
\end{equation}
In general, $h_{\phi}$ can depend on time $t$, $h_{\phi}\equiv
h_{\phi}(t)$. One meets this effective Hamiltonian when one starts
with the Schr\"{o}dinger Equation (\ref{Schrod}) for the total state
space ${\cal H}$ and looks for the rigorous evolution equation for
the distinguished subspace of states ${\cal H}_{||} \subset {\cal
H}$. In the case of one--dimensional ${\cal H}_{||}$  this rigorous
Schr\"{o}dinger--like evolution equation has the following form for
the initial condition $a(0) = 1$ (see \cite{PRA} and references one
finds therein),
\begin{equation}
i \hbar\, \frac{\partial a(t)}{\partial t} \;=\; h_{\phi}(t)\;a(t).
\label{eq-for-h}
\end{equation}
Relations (\ref{h}) and (\ref{eq-for-h}) establish a direct
connection between the amplitude $a(t)$ for the state $|\phi
\rangle$ and the exact effective Hamiltonian $h_{\phi}(t)$ governing
the time evolution in the one--dimensional subspace ${\cal H}_{\|}
\ni |\phi\rangle$. Thus the use of the evolution equation
(\ref{eq-for-h}) or the relation (\ref{h}) is one of the most
effective tools for the accurate analysis of the early-- as well as
the long--time properties of the energy and decay rate of a given
qausistationary state $|\phi (t) \rangle$.

So let us assume that we know the amplitude $a(t)$. Then starting
with this $a(t)$ and using the expression (\ref{h}) one can
calculate the effective Hamiltonian $h_{\phi}(t)$ in a general case
for every $t$. Thus, one finds the following expressions for the
energy and the decay rate of the system in the state $|\phi\rangle$
under considerations,
\begin{eqnarray}
{\cal E}_{\phi}&\equiv& {\cal E}_{\phi}(t) = \Re\,(h_{\phi}(t),
\label{E(t)}\\
\gamma_{\phi} &\equiv& \gamma_{\phi}(t) = -\,2\,\Im\,(h_{\phi}(t),
\label{G(t)}
\end{eqnarray}
where $\Re\,(z)$ and $\Im\,(z)$ denote the real and imaginary parts
of $z$ respectively.

As it was mentioned above the deviations of the decay law ${\cal
P}_{\phi}(t)$ from the exponential form can be described
equivalently using time-dependent decay rate. In terms of such
$\gamma_{\phi}(t)$ the Khalfin observation that ${\cal P}_{\phi}(t)$
must tend to zero as $t \rightarrow \infty$ more slowly than any
exponential function means that $\gamma_{\phi}(t) \ll
\gamma_{\phi}^{0}$ for $t \gg t_{as}$ and $\lim_{t \rightarrow
\infty} \,\gamma_{\phi}(t) = 0$.

Using (\ref{h}) and (\ref{E(t)}), (\ref{G(t)}) one can find that
\begin{eqnarray}
{\cal E}_{\phi} (0) &=& \langle \phi |H| \phi \rangle, \\
{\cal E}_{\phi} (t \sim \tau_{\phi}) & \simeq & {\cal E}_{\phi}^{0} \;\;\neq \;\; {\cal E}_{\phi} (0),\\
\gamma_{\phi}(0) &=& 0,\\
\gamma_{\phi}(t \sim \tau_{\phi}) &\simeq & \gamma_{\phi}^{0}.
\end{eqnarray}

The aim of this talk is to discuss the long time behaviour of ${\cal
E}_{\phi}(t)$  using $a(t)$ calculated for the given density
$\omega({\cal E})$. We show that ${\cal E}_{\phi}(t) \rightarrow {\cal E}_{min}$
as $t\rightarrow \infty$ for the model considered and that a wide
class of models has similar long time properties: ${{{\cal
E}_{\phi}(t)}\vline}_{\;t \rightarrow \infty} \neq {\cal
E}_{\phi}^{0}$. It seems that in contrast to the standard Khalfin
effect \cite{Khalfin} in the case of the quasistationary states
belonging to the same class as excited atomic levels, this long time
properties of the energy ${\cal E}_{\phi}(t)$ have a chance to be
detected, eg.,  by analyzing spectra of very distant stars.

\section{The model}

Let us assume that ${Spec. (H)} = [{\cal E}_{min}, \infty)$,
(where, ${\cal E}_{min} > - \infty$), and let us choose
$\omega ({\cal E})$ as follows
\begin{equation}
\omega ({\cal E}) = \frac{N}{2\pi}\,  \it\Theta ({\cal E} - {\cal E}_{min}) \
\frac{\gamma_{\phi}^{0}}{({\cal E}-{\cal E}_{\phi}^{0})^{2} +
(\frac{\gamma_{\phi}^{0}}{2})^{2}}, \label{omega-BW}
\end{equation}
where $N$ is a normalization constant and
\[
\it\Theta ({\cal E}) \ = \left\{
  \begin{array}{c}
   1 \;\;{\rm for}\;\; {\cal E} \geq 0, \\
   0 \;\; {\rm for}\;\; {\cal E} < 0,\\
  \end{array}
\right.
\]
For such $\omega ({\cal E})$ using (\ref{a-spec}) one has
\begin{equation}
a(t) = \frac{N}{2\pi}  \int_{{\cal E}_{min}}^{\infty}
 \frac{{\gamma_{\phi}^{0}}}{({\cal E}-{\cal E}_{\phi}^{0})^{2}
+ (\frac{\gamma_{\phi}^{0}}{2})^{2}}\, e^{\textstyle{ -
\frac{i}{\hbar}{\cal E}t}}\,d{\cal E}, \label{a-BW}
\end{equation}
where
\begin{equation}
\frac{1}{N} = \frac{1}{2\pi} \int_{{\cal E}_{min}}^{\infty}
 \frac{\gamma_{\phi}^{0}}{({\cal E}-{\cal E}_{\phi}^{0})^{2}
+ (\frac{\gamma_{\phi}^{0}}{2})^{2}}\, d{\cal E}. \label{N}
\end{equation}
Formula  (\ref{a-BW}) leads to the result
\begin{eqnarray}
a(t) &=& N\,e^{\textstyle{- \frac{i}{\hbar} ({\cal
E}_{\phi}^{0} -
i\frac{\gamma_{\phi}^{0}}{2})t}} \times \nonumber\\
&&\times \Big\{1 - \frac{i}{2\pi} \Big[
e^{\textstyle{\frac{\gamma_{\phi}^{0}t}{\hbar}}}\,
E_{1}\Big(-\frac{i}{\hbar}({\cal E}_{\phi}^{0} -{\cal E}_{min}
+ \frac{i}{2} \gamma_{\phi}^{0})t\Big) \nonumber\\
&&\;\;\;\;\;+ (-1) E_{1}\Big(- \frac{i}{\hbar}({\cal E}_{\phi}^{0} -{\cal E}_{min} -
\frac{i}{2} \gamma_{\phi}^{0})t\Big)\,\Big]\, \Big\}, \label{a-E(1)}
\end{eqnarray}
where $E_{1}(x)$ denotes the integral--exponential function
\cite{Sluis,Abramowitz}.

Using  (\ref{a-BW}) or (\ref{a-E(1)}) one easily finds that
\begin{equation}
i \hbar \,\frac{\partial a(t)}{\partial t} = ({\cal
E}_{\phi}^{0} - \frac{i}{2} \gamma_{\phi}^{0})\,a(t)\, +
\,\Delta a(t), \label{Delta -a}
\end{equation}
where
\begin{eqnarray}
\Delta a(t) &=& \frac{N}{\pi}\, \frac{\gamma_{\phi}^{0}}{2}
\,e^{\textstyle{- \frac{i}{\hbar}({\cal E}_{\phi}^{0} +
\frac{i}{2}\gamma_{\phi}^{0})t}}\times \nonumber \\
&& \;\;\;\;\;\;\;\;\times E_{1}\Big(-\frac{i}{\hbar}({\cal
E}_{\phi}^{0} - {\cal E}_{min} + \frac{i}{2}\, \gamma_{\phi}^{0})t \Big).
\label{Delta-a-E1}
\end{eqnarray}
So,
\begin{equation}
h_{\phi}(t) \equiv i \hbar \,\frac{\partial a(t)}{\partial
t}\,\frac{1}{a(t)} \stackrel{\rm def}{=} h_{\phi}^{0} +
\Delta h_{\phi}(t), \label{h+Delta-h}
\end{equation}
where
\begin{equation}
h_{\phi}^{0} \equiv {\cal E}_{\phi}^{0} - \frac{i}{2}\,
\gamma_{\phi}^{0}, \label{h-0}
\end{equation}
and
\begin{equation}
 \Delta h_{\phi}(t)  =  +
\,\frac{\Delta a(t)}{a(t)}. \label{Delta-h}
\end{equation}

Making use of  the asymptotic expansion of $E_{1}(x)$ \cite{Abramowitz},
\begin{equation}
{E_{1}(z)\vline}_{\, |z| \rightarrow \infty} \;\;\sim \;\;
\frac{e^{\textstyle{ -z}}}{z}\,( 1 - \frac{1}{z} + \frac{2}{z^{2}} -
\ldots ),  \label{E1-as}
\end{equation}
where $| \arg z  | < \frac{3}{2} \pi$, one finds
\begin{eqnarray}
{a(t)\vline}_{\, t \rightarrow \infty} &\simeq & N
e^{\textstyle - \frac{i}{\hbar}\,h_{\phi}^{0}\,t}
\nonumber \\&&
\;+\;
\frac{N}{2 \pi}\;e^{\textstyle{-\frac{i}{\hbar}\,{\cal E}_{min}t}}\;
\Big\{
(- i)\; \frac{
\gamma_{\phi}^{0}}{|\,h_{\phi}^{0}-{\cal E}_{min}\,|^{\,2}}
 \, \frac{\hbar}{t}  \nonumber \\
&&\;\;\;\;\;\;\;\;\;\;+ (- 2)\,\frac{({\cal E}_{\phi}^{0}\,-\,{\cal E}_{min})\,
\gamma_{\phi}^{0}}{|\,h_{\phi}^{0}\,-\,{\cal E}_{min}\,|^{\,4}} \,
\Big(\frac{\hbar}{t}\Big)^{2}\,+ \ldots\Big\} \label{a(t)-as}
\end{eqnarray}
and
\begin{eqnarray}
{\Delta a(t)\vline}_{\, t \rightarrow \infty} &\simeq &
\frac{N \gamma_{\phi}^{0}}{2 \pi}\;
e^{\textstyle{-\frac{i}{\hbar}\,{\cal E}_{min}t}}\;
\Big\{
\frac{h_{\phi}^{0}\,-\,{\cal E}_{min}}{|\,h_{\phi}^{0}\,-\,{\cal E}_{min}\,|^{\,2}} \; \frac{\hbar}{t} \nonumber \\
&&+ \;
\frac{(h_{\phi}^{0}\,-\,{\cal E}_{min})^{2}}{|\,h_{\phi}^{0}\,-\,{\cal E}_{min}\,|^{\,4}} \;
\Big(\frac{\hbar}{t}\Big)^{2}\;+\ldots \Big\}. \label{Da(t)-as}
\end{eqnarray}
These two last asymptotic expansions  enable one to find (see
(\ref{Delta-h}))
\begin{eqnarray}
{\Delta h_{\phi}(t)\vline_{\,t \rightarrow \infty}} & = &
{\frac{\Delta a(t)}{a(t)} \,\vline}_{\;t \rightarrow
\infty} \nonumber \\ &\simeq& - \, (h_{\phi}^{0}\,-\,{\cal E}_{min}\,)\,
-\,i\,\frac{\hbar}{t}\,  - \,2\, \frac{ {\cal E}_{\phi}^{0}\,-\,{\cal E}_{min}}{
|\,h_{\phi}^{0} \,-\,{\cal E}_{min}\,|^{\,2} } \Big( \frac{\hbar}{t} \Big)^{2}
+\ldots\;\;. \label{Delta-h-as}
\end{eqnarray}
Thus, there is
\begin{equation}
{h_{\phi}(t)\vline}_{\,t \rightarrow \infty} \,\simeq \,{\cal E}_{min}\,
-\,i\,\frac{\hbar}{t}\;  - \;2\, \frac{ {\cal E}_{\phi}^{0}\,-\,{\cal E}_{min}}{
|\,h_{\phi}^{0}\,-\,{\cal E}_{min} \,|^{\,2} }  \; \Big( \frac{\hbar}{t} \Big)^{2}
\;+\ldots \;\; \label{h-as}
\end{equation}
for the considered case (\ref{omega-BW}) of $\omega ({\cal E})$.

From (\ref{h-as}) it follows that
\begin{eqnarray}
\Re\,({h_{\phi}(t)\vline}_{\,t \rightarrow \infty}) \stackrel{\rm
def}{=} {\cal E}_{\phi}^{\infty}
\simeq  {\cal E}_{min}\, -\,2\,
\frac{ {\cal E}_{\phi}^{0}\,-\,{\cal E}_{min}}{ |\,h_{\phi}^{0}\,-\,{\cal E}_{min} \,|^{\,2} }  \; \Big(
\frac{\hbar}{t} \Big)^{2} \, \begin{array}{c}
                               {} \\
                               \longrightarrow \\
                               \scriptstyle{t \rightarrow \infty}
                             \end{array}
\, {\cal E}_{min},\label{Re-h-as}
\end{eqnarray}
where ${\cal E}_{\phi}^{\infty} = {\cal
E}_{\phi}(t)|_{\,t \rightarrow \infty}$,
and
\begin{equation}
\Im\,({h_{\phi}(t)\vline}_{\,t \rightarrow \infty}) \simeq
-\,\frac{\hbar}{t} \,
\begin{array}{c}
                               {} \\
                               \longrightarrow \\
                               \scriptstyle{t \rightarrow \infty}
                             \end{array}
\,0. \label{Im-h-as}
\end{equation}
The property (\ref{Re-h-as}) means that
\begin{equation}
\Re\,({h_{\phi}(t)\vline}_{\,t \rightarrow \infty})\,\equiv\,
{\cal E}_{\phi}^{\infty}\,
<\, {\cal E}_{\phi}^{0}. \label{E-infty<E0}
\end{equation}

For different states $|\phi\rangle \, =\,|j\rangle$, ($j = 1,2,3,\ldots$)
one has
\begin{eqnarray}
{\cal E}_{1}^{\infty} - {\cal E}_{2}^{\infty} &=&
-\,2\,\Big[ \frac{ {\cal E}_{1}^{0}\,-\,{\cal E}_{min}}{ |\,h_{1}^{0}\,-\,{\cal E}_{min}
\,|^{\,2} } \; -\; \frac{ {\cal E}_{2}^{0}\,-\,{\cal E}_{min}}{
|\,h_{2}^{0}\,-\,{\cal E}_{min} \,|^{\,2} } \;\Big]\; \Big( \frac{\hbar}{t}
\Big)^{2}\nonumber \\
&\neq& {\cal E}_{1}^{0} - {\cal E}_{2}^{0}\, \neq
0.\label{Re-h1-h2-as}
\end{eqnarray}
Note that
\begin{equation}
\Im\,({h_{1}(t)\vline}_{\,t \rightarrow \infty}) =
\Im\,({h_{2}(t)\vline}_{\,t \rightarrow
\infty}),\label{Im-h1-h2-as}
\end{equation}
whereas in general $\gamma_{1}^{0}\, \neq \,\gamma_{2}^{0}$.

The most interesting relations seems to be the following one
\begin{eqnarray}
\frac{{\cal E}_{1}^{\infty} - {\cal
E}_{2}^{\infty}}{{\cal E}_{3}^{\infty} - {\cal
E}_{\phi_{4}}^{\infty}} &=& \frac{\frac{{\cal
E}^{0}_{1}\,-\,{\cal E}_{min}}{|h^{0}_{1}\,-\,{\cal E}_{min}|^{2}} \,-\,\frac{{\cal
E}^{0}_{2}\,-\,{\cal E}_{min}}{|h^{0}_{2}\,-\,{\cal E}_{min}|^{2}}}{\frac{{\cal
E}^{0}_{3}\,-\,{\cal E}_{min}}{|h^{0}_{3}\,-\,{\cal E}_{min}|^{2}} \,-\,\frac{{\cal
E}^{0}_{4}\,-\,{\cal E}_{min}}{|h^{0}_{4}\,-\,{\cal E}_{min}|^{2}}}
\label{E1-E2|E3-E4}\\
& \neq & \frac{{\cal E}_{1}^{0} - {\cal
E}_{2}^{0}}{{\cal E}_{3}^{0} - {\cal
E}_{4}^{0}}, \nonumber
\end{eqnarray}
and
\begin{equation}
\frac{\Im\,({h_{1}(t)\vline}_{\,t \rightarrow
\infty})}{\Im\,({h_{2}(t)\vline}_{\,t \rightarrow
\infty})}\;=\;1\;\neq\;\frac{\gamma_{1}^{0}}{\gamma_{2}^{0}}.
\label{Im-h1|Im-h2}
\end{equation}
The relation (\ref{E1-E2|E3-E4}) is valid also when one takes $|1\rangle$ instead of $|3\rangle$ or
$|2\rangle$ instead of $|4\rangle$.
It seems  to be interesting that relations (\ref{E1-E2|E3-E4}),
(\ref{Im-h1|Im-h2}) do not depend on time $t$.

Note that the following conclusion can be drawn from
(\ref{Re-h1-h2-as}):  For suitably long times $t > t_{as}$ there must be
\begin{equation}
\vline \; {\cal E}_{1}^{\infty}\, -\, {\cal
E}_{2}^{\infty}\,\vline \;\; <\;\; \vline\,{\cal
E}_{1}^{0} - {\cal E}_{2}^{0}\vline \,.
\label{E1-E2-as}
\end{equation}

These suitable times can be estimated using relation
(\ref{a(t)-as}). From (\ref{a(t)-as}) one obtains
\begin{equation}
{\vline\,{a(t)\vline}_{\, t \rightarrow
\infty}\,\vline}^{\,2} \simeq N^{2} e^{\textstyle - \frac{
\gamma_{\phi}^{0}}{\hbar}\,\,t}\; +\; \frac{N^{2}}{4
\pi^{2}}\;\frac{(\gamma_{\phi}^{0})^{2}}{|h_{\phi}^{0}\,-\,{\cal E}_{min}\,|^{\,4}} \;
\frac{\hbar^{2}}{t^{2}}\; + \;\ldots\;\; . \label{t-as-1}
\end{equation}
Relations (\ref{Delta-h-as}) --- (\ref{E1-E2-as}) become important
for times $t > t_{as}$, where $t_{as}$ denotes the time $t$ at which
contributions to ${\vline\,{a(t)\vline}_{\, t \rightarrow
\infty}\,\vline}^{\,2}$ from the first exponential component in
(\ref{t-as-1}) and from the second component proportional to
$\frac{1}{t^{2}}$ are comparable. So $t_{as}$ can be be found by
considering the following relation
\begin{equation}
e^{\textstyle - \frac{ \gamma_{\phi}^{0}}{\hbar}\,\,t}\; \sim\;
\frac{1}{4
\pi^{2}}\;\frac{(\gamma_{\phi}^{0})^{2}}{|h_{\phi}^{0}\,-\,{\cal E}_{min}\,|^{\,4}} \;
\frac{\hbar^{2}}{t^{2}}. \label{t-as-2}
\end{equation}
Assuming that the right hand side is equal to the left hand side in
the above relation one gets a transcendental equation.  Exact
solutions of such an equation can be expressed by means of the
Lambert $W$ function \cite{Corless}. An asymptotic solution of the
equation obtained from the relation (\ref{t-as-2}) is relatively
easy to find \cite{Olver}. The very approximate asymptotic solution,
$t_{as}$, of this equation for $(\frac{{\cal
E}_{\phi}}{\gamma_{\phi}^{0}})\,>\,10^{\,2}$ (in general for
$(\frac{{\cal E}_{\phi}}{\gamma_{\phi}^{0}})\,\rightarrow \,\infty$)
has the form
\begin{eqnarray}
\frac{\gamma_{\phi}^{0}\,t_{as}}{\hbar} &\simeq & 8,28 \,+\, 4\,
\ln\,(\frac{{\cal E}_{\phi}^{0}\,-\,{\cal E}_{min}}{\gamma_{\phi}^{0}}) \nonumber \\
&&+\, 2\,\ln\,[8,28 \,+\,4\,\ln\,(\frac{{\cal
E}_{\phi}^{0}\,-\,{\cal E}_{min}}{\gamma_{\phi}^{0}})\,]\,+\, \ldots \;\;.
\label{t-as-3}
\end{eqnarray}

\section{Some generalizations}

To complete the analysis performed in the previous Section let us
consider a more general case of $a(t)$. Namely, let the
asymptotic approximation to $a(t)$ have the form
\begin{equation}
a(t) \;\;
\begin{array}{c}
   {} \\
   \sim\\
   {\scriptstyle t \rightarrow \infty}
 \end{array}
\;\;e^{\textstyle{-i\frac{t}{\hbar}\,{\cal E}_{min}}}\; \sum_{k=0}^{N} \,\frac{c_{k}}{t^{\lambda + k}},
\label{a(t)-as-w}
\end{equation}
where $\lambda > 0$ and $c_{k}$ are complex numbers. The simplest case occurs for ${\cal E}_{min} = 0$.
Note that the asymptotic expansion for $a(t)$ of this or a similar form
one obtains for a wide class of densities of energy distribution
$\omega({\cal E})$ \cite{Khalfin,Goldberger,Fonda,Peres,Arbo},
\cite{seke} -- \cite{jiitoh}, \cite{Dittes}.

From the relation (\ref{a(t)-as-w}) 
one concludes that
\begin{equation}
\frac{\partial a(t)}{\partial t} \;\;
\begin{array}{c}
   {} \\
   \sim\\
   {\scriptstyle t \rightarrow \infty}
 \end{array}
\;\; e^{\textstyle{-i\frac{t}{\hbar}\,{\cal E}_{min}}}\;\Big\{-\frac{i}{\hbar}\, {\cal E}_{min}\,
-\,\sum_{k=0}^{N} \,(\lambda + k)\,\frac{c_{k}}{t^{\lambda + k
+ 1}}\Big\}. \label{da(t)-as-w}
\end{equation}

Now let us take into account the relation (\ref{eq-for-h}). From
this relation and relations (\ref{a(t)-as-w}), (\ref{da(t)-as-w}) it
follows that
\begin{equation}
h_{\phi}(t) \;\;
\begin{array}{c}
   {} \\
   \sim\\
   {\scriptstyle t \rightarrow \infty}
 \end{array}
\;\;
{\cal E}_{min}\,+\,
\frac{d_{1}}{t}\, +
\,\frac{d_{2}}{t^{2}}\,+\,\frac{d_{3}}{t^{3}}\,+\,\ldots\,\,\, ,
\label{h-sim-w}
\end{equation}
where $d_{1}, d_{2}, d_{3}, \ldots$ are complex numbers with negative or positive real and imaginary parts. This means
that in the case of the asymptotic approximation to $a(t)$ of
the form (\ref{a(t)-as-w}) the following  property holds,
\begin{equation}
\lim_{t \rightarrow \infty}\,h_{\phi}(t) \, = \, {\cal E}_{min}\, <\, {\cal E}_{\phi}^{0}. \label{lim-h}
\end{equation}

It seems to be important that  results (\ref{h-sim-w}) and
(\ref{lim-h}) coincide with the results (\ref{h-as}) ---
(\ref{Re-h1-h2-as}) obtained for the density $\omega ({\cal E})$
given by the formula (\ref{omega-BW}). This means that general
conclusion obtained for the other $\omega ({\cal E})$ defining
unstable states should be similar to those following from
(\ref{h-as}) --- (\ref{Re-h1-h2-as}).

\section{Final remarks.}

Let us consider a class of unstable states formed by excited atomic
energy levels and let these excited atoms emit the
electromagnetic waves of the energies ${\cal E}_{\phi}^{0} = {\cal E}_{n_{jk}}^{0} \equiv
h\,\nu_{n_{jk}}^{0}$ (where $\nu_{n_{jk}}^{0}$ denotes the frequency
of the emitted wave, and ${\cal E}_{n_{jk}}^{0}$ is
energy emitted by an electron jumping
from the energy level $n_{j}$ to  the energy level $n_{k}$). Then for times $ t > t_{as}$,
according to the results of the previous
sections, there should be
\begin{equation}
{\cal E}_{n_{jk}}^{\infty}\,<\, {\cal E}_{n_{jk}}^{0}. \label{E-infty-E-0}
\end{equation}
So in the case of electromagnetic radiation in the optical range
registered by a suitably  distant observer this effect should manifest
itself as a red shift. In a general case this effect should cause a loss of the
energy of the emitted electromagnetic radiation if the distance between an emiter
and receiver is suitably long, that is if the emitted radiation reaches such a distance from
the emitter that the time necessary for photons to reach this distance is longer than
a maximal range of time of the validity of the exponential decay law for the excited atomic
level emitting this radiation.

It can be easily verified that  relation (\ref{E1-E2|E3-E4}) do not depend on the redshift
connected with the Doppler effect \cite{Heitler}.
Indeed let the excited atoms emit the
electromagnetic waves
and let them move away from the observer with
the velocity $v$. Then, the observer's measuring devices will show
that the energies emitted by these moving atoms are ${\cal
E}_{n_{jk}}^{0,\,v}$. The energy ${\cal E}_{n_{jk}}^{0}$ emitted by
this same source at rest is connected with ${\cal
E}_{n_{jk}}^{0,\,v}$ by the Doppler formula \cite{Heitler}
\begin{equation}
{\cal E}_{n_{jk}}^{0,\,v} = \kappa\;{\cal E}_{n_{jk}}^{0}\; <
\;{\cal E}_{n_{jk}}^{0}, \label{E-v}
\end{equation}
where $ \kappa =  \frac{1 - \beta}{\sqrt{1 - \beta^{2}}}$ and $\beta
= \frac{v}{c}$, and
\begin{equation}
{\cal E}_{n_{1k}}^{0,\,v} - {\cal E}_{n_{2k}}^{0,\,v} = \kappa\,
({\cal E}_{n_{1k}}^{0} - {\cal E}_{n_{2k}}^{0}). \label{E1-E2-v}
\end{equation}
From (\ref{E-v}) and (\ref{E1-E2-v}) it follows that
\begin{equation}
\frac{{\cal E}_{n_{1k}}^{0,\,v} - {\cal E}_{n_{2k}}^{0,\,v}}{{\cal
E}_{n_{3k}}^{0,\,v} - {\cal E}_{n_{4k}}^{0,\,v}} \,=\, \frac{{\cal
E}_{n_{1k}}^{0} - {\cal E}_{n_{2k}}^{0}}{{\cal E}_{n_{3k}}^{0} -
{\cal E}_{n_{4k}}^{0}}. \label{E1-E2|E3-E4-Dop}
\end{equation}

Now, let us assume for a moment that the $\omega({\cal E})$ given by
the formula (\ref{omega-BW}) describes the considered sources with
the sufficient accuracy.  Next if one additionally assumes that in
the distant past photons of energies, say ${\cal E}_{n_{jk}}^{0}$,
($j=1,2$), were emitted and that the sources of this emission were
moving away from the observer with known constant velocity $v$, then
at the present epoch this observer detects energies ${{\cal
E}_{n_{jk}}^{\infty\;}}' \stackrel{\rm def}{=} \kappa\, {\cal E}_{n_{jk}}^{\infty}$,
where ${\cal E}_{n_{jk}}^{\infty} \stackrel{\rm def}{=}
\Re\,({h_{n_{jk}}(t)\vline}_{\,t \rightarrow \infty}$. If the moment
of this emission was at suitably  distant past then according to the
results of Sec. 2 (see (\ref{E1-E2-as})) it should appear that
\begin{equation}
|{{\cal E}_{n_{1k}}^{\infty\;}}'\, -\, {{\cal
E}_{n_{2k}}^{\infty\;}}'\,|\;<\;|\, {{\cal E}_{n_{1k}}^{0,\,v}} -
{{\cal E}_{n_{2k}}^{0,\,v}}\,|\,= \,\kappa\,|\, {{\cal E}_{n_{1k}}^{0}} -
{{\cal E}_{n_{2k}}^{0}}\,|, \label{E1-E2-infty}
\end{equation}
and
\begin{equation}
\frac{{\cal E}_{n_{1k}}^{\infty} - {\cal E}_{n_{2k}}^{\infty}}{{\cal
E}_{n_{3k}}^{\infty} - {\cal E}_{n_{4k}}^{\infty}} \,\neq\,
\frac{{\cal E}_{n_{1k}}^{0} - {\cal E}_{n_{2k}}^{0}}{{\cal
E}_{n_{3k}}^{0} - {\cal E}_{n_{4k}}^{0}}. \label{E1-E2|E3-E4-infty}
\end{equation}
Therefore
it seems that there is a chance to detect the possible effect described in this paper
using relation  (\ref{E1-E2|E3-E4}) and analyzing spectra of distant astrophysical objects.
It can be done using this relation if one is able to register and analyze at least three
different emission  lines from the same distant source.
Another possibility  to observe this effect is to modify the experiment described in \cite{rothe}
in such a way that the emitted energy (frequency) of the luminescence decays could be measured
which could make possible to test  relations (\ref{E-infty-E-0}) or (\ref{E1-E2-as}).

Unfortunately, the model considered in Sec. 2 does not reflect
correctly all properties of the real physical system containing
unstable states. Therefore, formula (\ref{E1-E2|E3-E4}) obtained
within this model can not be considered as universally valid and
it does not lead to the expressions  correctly describing real properties of very
distant astronomical sources of electromagnetic radiation. The
defect of this model is that some quantities calculated within it
are divergent. Indeed from (\ref{a(t)}), (\ref{a-spec}) and
(\ref{Schrod}) one finds that
\begin{equation}
i \hbar\,{{\frac{\partial a(t)}{\partial t}}\vline}_{\,t=0} =
\langle\phi|H|\phi\rangle, \label{da|dt-t=0}
\end{equation}
but inserting into (\ref{a-spec}) the density $\omega ({\cal E})$
given by (\ref{omega-BW}),  relation (\ref{da|dt-t=0}) yields
$\langle\phi|H|\phi\rangle = \infty$. For this same reason
 $h_{\phi}(t)$ and $\Delta h_{\phi}(t)$ computed for
$\omega ({\cal E})$ given by (\ref{omega-BW}) are divergent at
$t=0$: $ h_{\phi}(=0) = \infty, \;\; \Delta h_{\phi}(t=0) = \infty
$. Nevertheless, taking into account that a very wide class of
models of unstable states leads to a similar asymptotic long time
behaviour of $a(t)$ to this obtained for the model considered
in Sec. 2, some general conclusions following from the results of
Sec. 2 and 3 seem to deserve more attention.

\end{document}